\documentclass[preprint,journal]{vgtc}       

\ifpdf
  \pdfoutput=1\relax                   
  \usepackage{graphicx}                
  \DeclareGraphicsExtensions{.pdf,.png,.jpg,.jpeg} 
\else
  \usepackage{graphicx}                
  \DeclareGraphicsExtensions{.eps}     
\fi%
\usepackage{todonotes}
\graphicspath{{figures/}{}{images/}{./}} 

\usepackage{microtype}                 
\PassOptionsToPackage{warn}{textcomp}  
\usepackage{textcomp}                  
\usepackage{mathptmx}                  
\usepackage{times}                     
\usepackage{cite}                      
\usepackage{tabu}                      
\usepackage{booktabs}                  

\newcommand{\cmmnt}[1]{}

\usepackage{xcolor}
\definecolor{bronze}{rgb}{0.8, 0.5, 0.2}

\definecolor{mygray}{rgb}{0.66, 0.66, 0.66}

\ieeedoi{10.1109/TVCG.2019.2934260}

\vgtccategory{Research}
\vgtcpapertype{technique}

\title{Artifact-Based Rendering: Harnessing Natural and Traditional Visual Media for More Expressive and Engaging 3D Visualizations}

\author{Seth Johnson, \textit{Student Member, IEEE}, Francesca Samsel, Gregory Abram, Daniel Olson, Andrew J. Solis,\\
Bridger Herman, Phillip J. Wolfram, Christophe Lenglet, Daniel F. Keefe, \textit{Member, IEEE}}
\authorfooter{
\item
 Seth Johnson, Daniel Olson, Bridger Herman, Christophe Lenglet, and Daniel F. Keefe are with the University of Minnesota.  E-mails: \{joh08230,olso7118,herma582,clenglet,dfk\}@umn.edu.
\item
 Francesca Samsel and Gregory Abram are the the University of Texas at Austin.  E-mails: fsamsel@tacc.utexas.edu, gda@tacc.utexas.edu.
\item
 Phillip Wolfram is with the Fluid Dynamics and Solid Mechanics, Theoretical Division, Los Alamos National Laboratory. E-mail: pwolfram@lanl.gov. 
}

\shortauthortitle{Johnson \MakeLowercase{\textit{et al.}}: Artifact-Based Rendering}

\abstract{We introduce Artifact-Based Rendering (ABR), a framework of tools, algorithms, and processes that makes it possible to produce real, data-driven 3D scientific visualizations with a visual language derived entirely from colors, lines, textures, and forms created using traditional physical media or found in nature.  A theory and process for ABR is presented to address three current needs:  (i) designing better visualizations by making it possible for non-programmers to rapidly design and critique many alternative data-to-visual mappings; (ii) expanding the visual vocabulary used in scientific visualizations to depict increasingly complex multivariate data; (iii) bringing a more engaging, natural, and human-relatable handcrafted aesthetic to data visualization.  New tools and algorithms to support ABR include front-end applets for constructing artifact-based colormaps, optimizing 3D scanned meshes for use in data visualization, and synthesizing textures from artifacts.  These are complemented by an interactive rendering engine with custom algorithms and interfaces that demonstrate multiple new visual styles for depicting point, line, surface, and volume data.  A within-the-research-team design study provides early evidence of the shift in visualization design processes that ABR is believed to enable when compared to traditional scientific visualization systems.  Qualitative user feedback on applications to climate science and brain imaging support the utility of ABR for scientific discovery and public communication.
} 

\keywords{Visualization Design, Art and Visualization, Data Physicalization, Multivariate Visualization.}

\CCScatlist{ 
 \CCScat{K.6.1}{Management of Computing and Information Systems}%
{Project and People Management}{Life Cycle};
 \CCScat{K.7.m}{The Computing Profession}{Miscellaneous}{Ethics}
}

\teaser{
\centering
\includegraphics[width=\linewidth]{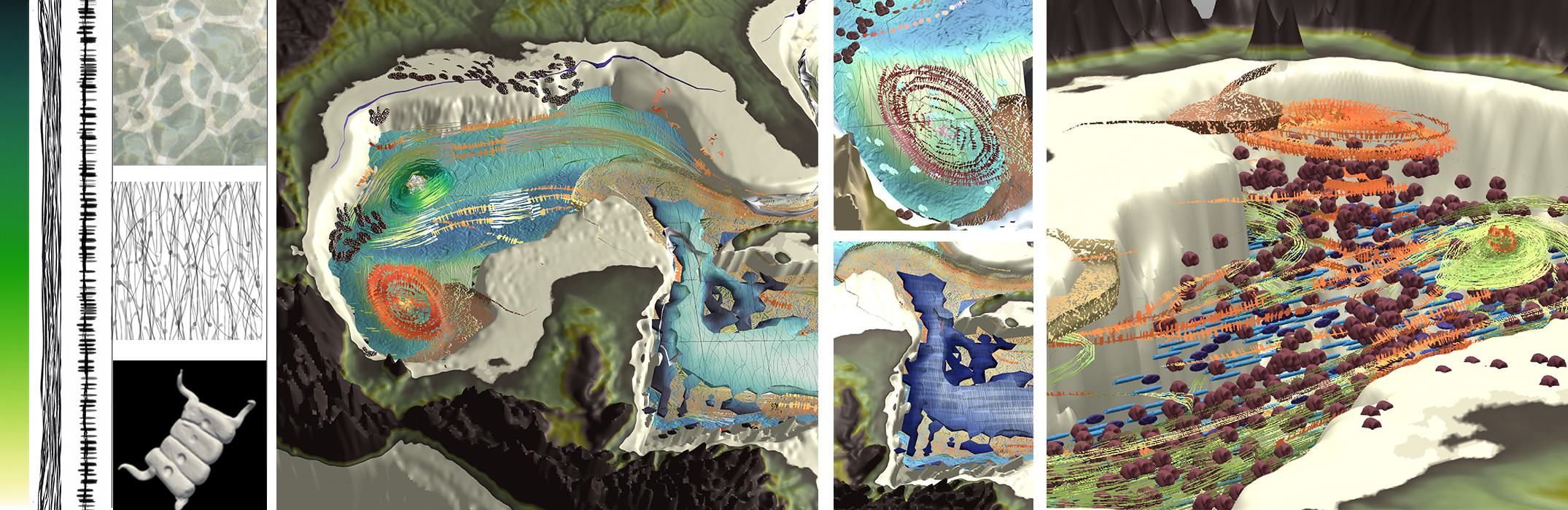}
\caption{Using traditional physical artistic media as input to the digital visualization pipeline provides a richer visual vocabulary and opens the door for artists to participate in creating more expressive and engaging 3D scientific visualizations.  This example helps scientists understand commercially viable macroalgae growth in the Gulf of Mexico by encoding temperature and salinity from remote sensing together with eddy direction and curvature and three nitrate concentrations from computational simulation.}
\label{fig:teaser}
}

\vgtcinsertpkg

\begin{document}


\firstsection{Introduction}

\maketitle

Finding inspiration in our physical world, combining disparate objects in new ways, understanding through hands-on making --- for centuries these ``low-tech'', physical processes have helped us (scientists, artists, architects, doctors, engineers) to investigate, reinterpret, and ultimately make sense of our world.  Visualization, particularly in immersive environments such as virtual or augmented reality (VR or AR), promises a similar, more physical, perhaps even innately human approach to making sense of today's complex data.  Yet, current computer-based visualizations fall short of realizing many of the benefits of more traditional, time-tested, physical sense-making processes.

We highlight three specific challenge problems for the future of scientific visualization:

\textbf{Challenge 1.}  Supporting traditional visual designers and design processes, such as those taught in art and design disciplines.  Here,``supporting'' means, in part, creating new visualization design tools that make it possible for non-programmers to rapidly design and critique many alternative data-to-visual mappings.  

\textbf{Challenge 2.}  Expanding the visual vocabulary used in  visualizations in order to depict increasingly complex multivariate data.  The consistent computer-generated aesthetic found throughout our conference proceedings is becoming well refined, but it is in sharp contrast to the much larger visual variety we find when walking the halls of a museum or even walking through the woods.  Is it possible that our visual vocabulary is converging upon a local, rather than global, maximum?  We ask, to what extent might a richer visual vocabulary increase overall expressiveness, with new visual encodings or new combinations of encodings conveying additional or more complex data?

\textbf{Challenge 3.}  Bringing a more engaging, natural, and human-relatable handcrafted aesthetic to data visualization.  We live in a time when there is a great disconnect between the scientific community and ``lay people''.  Scientists struggle to tell their stories.  In theory, visualization should be the most powerful tool scientists have to communicate with each other and the public, but the disconnect persists.  Thus, we ask, how might even the methods we choose to depict scientific data reaffirm the natural, human connection to all aspects of science?

In formulating Challenge 3, we have closely followed recent research in data physicalization~\cite{alexander2015exploring,taylor2015data,khot2014understanding,djavaherpour2017physical}.  Benefits of data representations that are physical, real-world objects rather than purely digital representations include the potential to use multi-sensory perception and dual encodings to increase data legibility~\cite{hogan2016towards}. The potential benefits also include increased engagement and emotional connection with data~\cite{jansen2015opportunities}.  Imagine touching a physical, data-driven melting ice sculpture depicting in physical form how the terminus of Grewingk Glacier has receded over 150 years~\cite{sengal2015glacier}, and compare this to a typical online map-based visualization of the same data.  Certainly a map will be better for some  data analysis tasks, but the scientific community cannot dismiss the sculptural visualization.  Cognitive science demonstrates that humans are innately compelled to touch and examine some physical objects~\cite{klatzky2012please}.  We engage and connect with natural, physical forms, especially when there is evidence of the human hand in them.  Thus, a physical ice sculpture can be just as valuable to science and society as a finely crafted digital map and, perhaps, more impactful in the actions it inspires.

Our work seeks to extend physicalization techniques by introducing an inverse problem relative to what has been studied thus far, namely, using physicalization as an input to the visualization pipeline rather than just an output.  We reason that the resulting, often handcrafted, aesthetic can have many benefits, including highlighting the human connection to the data.  In a sense, this approach could mimic for 3D scientific data the hand-drawn 2D aesthetic Georgia Lupi has leveraged so effectively to advance her work on ``data humanism''~\cite{lupi2017datahumanism}.

Our work also closely follows research in visualization design tools, particularly those supporting artists' contribution to scientific visualization, which inspires Challenges 1 and 2.  In the tradition of the ``renaissance teams'' introduced by Donna Cox~\cite{cox1988using}, our interdisciplinary team includes computer scientists, domain scientists, and a traditionally trained artist.  Without new tools, it has been impossible to translate the multitude of stunning visual design ideas developed by the artist using traditional media (e.g., paper, ink, clay, wax) into 3D data-driven visualizations.  Prior software systems and design processes have identified and addressed this same problem but incompletely.  Scientific Sketching supports artists prototyping 3D visuliazations in VR, but not in a data-driven way~\cite{keefe2008scientific}.  Vis-by-Sketching supports artists in data-driven prototyping, but only for 2D spatial datasets~\cite{schroeder2016visualization}.  Our work is the next logical progression, extending to support rapid data-driven 3D visualization design and prototyping.

The major novel idea in this paper is, therefore, to introduce a decidedly human, physical approach to crafting data visualizations by working with artists and the traditional media that they can so powerfully control (Fig.~\ref{fig:teaser}).  We demonstrate how all major visual elements of 3D multivariate scientific visualizations (color, line, texture, and form) can be derived from physical artifacts designed and crafted or even found in nature by these artists.  Further, with a custom rendering engine, the artifacts can respond dynamically to data to produce complete, accurate, data-driven interactive visualizations.  We call the framework of tools, algorithms, and processes that comprise this idea Artifact-Based Rendering (ABR). 

The framework serves as a visualization design tool, but, to reiterate, the goal is not as simple as producing only efficient, effective solutions.  A key anticipated benefit of ABR is opening the field to a more diverse group of visual designers who might contribute radically more effective solutions that have not yet been discovered.  Since the foundation of creativity and design in traditional artistic fields is rapid exploration of alternatives (e.g., think sketching~\cite{buxton2010sketching}); it follows that a key requirement for the framework is to support rapid exploration of many alternatives.  ABR does this by leveraging the skills artists have with traditional media.  While this may not be the most efficient solution for computer technologists to design visualizations, artists can create dozens of novel 3D glyph designs in clay in just one hour.

As they work, our experience is that artists will often naturally follow visualization design guidelines established through other means (e.g., perceptual studies~\cite{Ware2012infoVis}), many of these are, after all, inspired by classic color and design theory.  However, artists are sure to also occasionally break these rules, and this can be useful too, for example, by combining texture and glyphs in a new way inspired by traditional sculpture or by helping us craft custom visualizations that are intentionally not generalizable but instead address a new goal of optimally conveying specific scientific results to the public.

From a technical standpoint, several computing advances were required in order to present a first implementation of ABR.  For example, we combine 2D and 3D scanning techniques, texture synthesis, and morphing in new ways for visualization while also supporting interactive 3D rendering.  The main contributions of this paper can be summarized as:
(1) The concept and theory of Artifact-Based Rendering.
(2) The design of four example front-end applets that prepare artifacts for visualization by constructing colormaps, cropping and calculating normal maps from image artifacts, synthesizing new textures from examples, and optimizing 3D scanned meshes.
(3) The specification for and implementation of a first ABR rendering engine with custom algorithms and interfaces to enable multiple new visual styles for depicting point, line, surface, and volume data. 
(4) Evaluation of impact on visualization design processes via a within-the-research-team design study comparing designing visualizations with ABR versus a traditional tool.
(5) Results and feedback for two domain science applications.

\section{Related Work}
\subsection{Artistic Techniques and Theories in Visualization}
The field of data visualization has often benefited from art and design theories, processes, and  techniques.  Examples include accentuating the legibility of 3D forms by rendering them in the style of pen and ink illustration~\cite{winkenbach1994computer} or with shading based on artistic color theory~\cite{gooch1998non}.  Inspired by traditional oil painting ~\cite{kirby2005painting}, researchers have also developed algorithms to render data-driven ``brushstrokes'' for visualization that might translate to an improved ability to support multi-level understanding of data.  Some results utilize painterly layering and composition but maintain a mostly geometric appearance~\cite{kirby1999visualizing,laidlaw1998visualizing}.  In others, brushstrokes are clearly visible~\cite{healey2002perception,healey2001formalizing,tateosian2007engaging}.  These look ``painted'' to most of us; however, artists critique the visualizations as regular and algorithmic, missing the richness and subtlety of a traditional painting.  

Our work advances this research on three fronts.  First, we extend to 3D the core concept of building a visualization up from data-driven, artistic, low-level visual elements.  Second, we introduce a method for accomplishing this with real, physical artifacts rather than via algorithmically generated approximations, which are often limited in their ability to capture the original artistic intent.  Finally, we include actual artists in the process of crafting these visualizations.

\subsection{Artists and Designers in Visualization}
The hand-drawn ``Dear Data'' series of 2D information visualizations~\cite{deardata}, demonstrate how artists themselves can develop their own visual languages to convey data in ways that are both accurate and inspire human connection to the underlying information.  These results build upon a tradition of artists' contributing to visualization (e.g., ~\cite{cox1988using}) and advance goals of the National Academies of Sciences, Engineering, and Medicine, which call for expanded art-science collaborations~\cite{SEAD}.  

New tools are required to support artist involvement~\cite{kahler2002rendering}, and researchers have developed several.  Drawing with the Flow~\cite{schroeder2010drawing} and later Visualization-by-Sketching~\cite{schroeder2016visualization} do this with custom pen-based user interfaces.  Visualization-by-Sketching supports multivariate data layers and animated streaklets, but all in 2D.  Volume Shop~\cite{bruckner2005volumeshop} and WYSIWYG Volume Rendering~\cite{guo2011wysiwyg} make designing transfer functions for volume renderings accessible to artists, but this does not quite address the multivariate design challenge in the sense that volume rendering is hard to extend beyond conveying 1 or 2 variables simultaneously.  Scientific Sketching~\cite{keefe2005artistic,keefe2008scientific} is something like a VR sketchbook for artists to design scientific visualizations.  So, it is 3D and expressive, but it does not connect to any underlying data in order to turn the sketches into real, data-driven visualizations.  Our work is the first in which artists may construct 3D data-driven visualizations using visual elements they craft themselves using the traditional media with which they are already experts.

\subsection{Data Physicalization and Human Connection}
Our approach is closely related to emerging research in ``data physicalization''~\cite{jansen2015opportunities}.  Whereas data physicalization is the process of visualizing data via a physical output (3D printouts, sculptures, active touch tables, etc.), ABR is the inverse, using curated or handcrafted physical objects as inputs to generate digital data visualizations.  In concept, this builds upon the work of artists, such as Mielbach~\cite{fs_spectrum}, who use physical materials to provide context and connection for science.  Both visual artists and psychologists have studied the geometric characteristics of shape (e.g., roundness, angularity, simplicity, complexity) and their impacts on human emotional responses~\cite{Lu:2012ip}.  There is evidence that data visualizations can be more effective and engaging when created by hand.  Route maps can be more effective when presented in a hand-drawn style~\cite{agrawala2001rendering}, and hand-draw iconography improves engagement and retention in data-driven storytelling~\cite{lee2013sketchstory,discoveryjam2}.

Glyphs (see surveys \cite{Ward:2002:taxonomy,fuchs2017systematic,ropinski2011survey}) are one area where we believe ABR can make a powerful contribution to visualization.  Prior glyph designs are characteristically geometric in their visual aesthetic. This is true for glyphs formed by superposition of 3D primitives (cones, spheres, cubes)~\cite{feng2009evaluation,GlyphToolbox,Lombeyda16dis3Dg,Legg_glyph_failsafe} or parametric surfaces that are elegantly defined to vary in response to multiple data axes~\cite{kindlmann2004superquadric}.  
In contrast, our collaborative project grew out of discussions of how an artist would approach designing similar glyphs.  The discussion quickly turned to a demonstration, when a box of 250 small handcrafted clay glyph ``sculpturettes'' arrived by mail from the artist.  With a bit of clay in hand, a sculptor can create 40 to 50 alternative designs for 3D glyphs that could be used for multivariate flow visualization within an hour.  

\subsection{Colormaps and Textures for Visualization}

Seminal research on designing and testing perceptually accurate colormaps for general use in visualizations \cite{Ware2012infoVis,moreland_diverging_2009,rogowitzWhich2001,ZhouSurvey,rheingans2000task} establishes several rules of thumb, such as relying primarily on luminance and saturation for depicting magnitude data.  Tools are also available for evaluating and modifying maps to adhere to commonly accepted guidelines \cite{Bujack,kovesi_good_2014,ZhouSurvey}, which closely parallel the fundamentals of color theory as studied by artists \cite{albers_interaction_2009,itten_elements_1970}. Our work shares a similar motivation with systems that leverage artistic color theory, example works of art, or artists themselves to design effective colormaps.  Our Color Loom applet extracts the color palette automatically from source images, which can be works of art, similar to manual approach introduced by Vote et al.~\cite{Vote-2003-DBE}.  Because they are quick to create, and artists can tune the results based on the data, the resulting colormaps are often useful for revealing more information in specific datasets \cite{Samsel2016asteroidvis,samselCHI2015} or even better engaging users, as in recent studies of affective use of color~\cite{Samsel_Affect}.  

Although not as common as colormapping, varying texture in response to data is a technique that has also been used previously~\cite{ware1995using, healey1999large, laidlaw1998visualizing}, including to encode uncertainty~\cite{botchen2005texture}.  Closely related to our work is that of Interrante et al. who encoded data using natural 2D textures of fibers and weavings of different densities~\cite{interrante2000harnessing}.  Later Gorla et al. extended this to synthesize texture from an example that follows a vector field on a 3D surface~\cite{gorla2003texture}.  We identify data-driven synthesis as important step for future work.


\begin{figure}[tb]
\centering
\includegraphics[width=\linewidth]{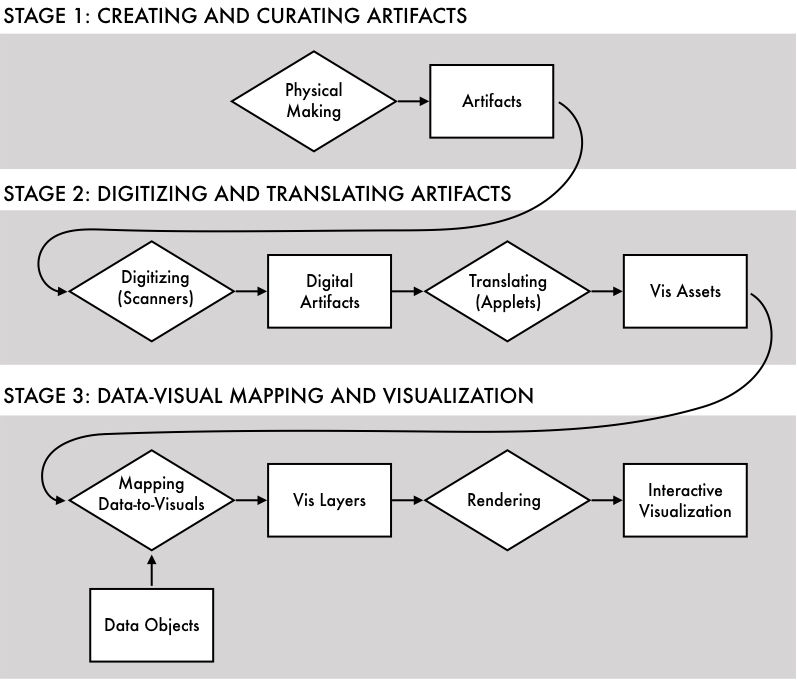}
\caption{The ABR pipeline contains three main stages.}
\label{fig:pipeline}
\end{figure}

\section{Artifact-Based Rendering for Visualization}
ABR is a framework of tools, algorithms, and processes that makes it possible to produce real, data-driven 3D scientific visualizations with a visual language derived entirely from colors, lines, textures, and forms created using traditional physical media or found in nature.  This section presents the ABR theory, processes, and technical system details that have been developed through a two-year, iterative  process.

Fig. 2 diagrams the full pipeline for ABR  visualization, divided into three stages: (1) Physical design work to craft artifacts; (2) Digitizing and translating artifacts for data-driven visualization; (3) Creating data-to-visual mappings to implement multivariate interactive visualizations.  Given the current library of pre-loaded artifacts, it is possible to begin a new project at any stage and then return to earlier stages as needed to create or adjust artifacts.

\begin{figure}[tb]
\centering
\includegraphics[width=\columnwidth]{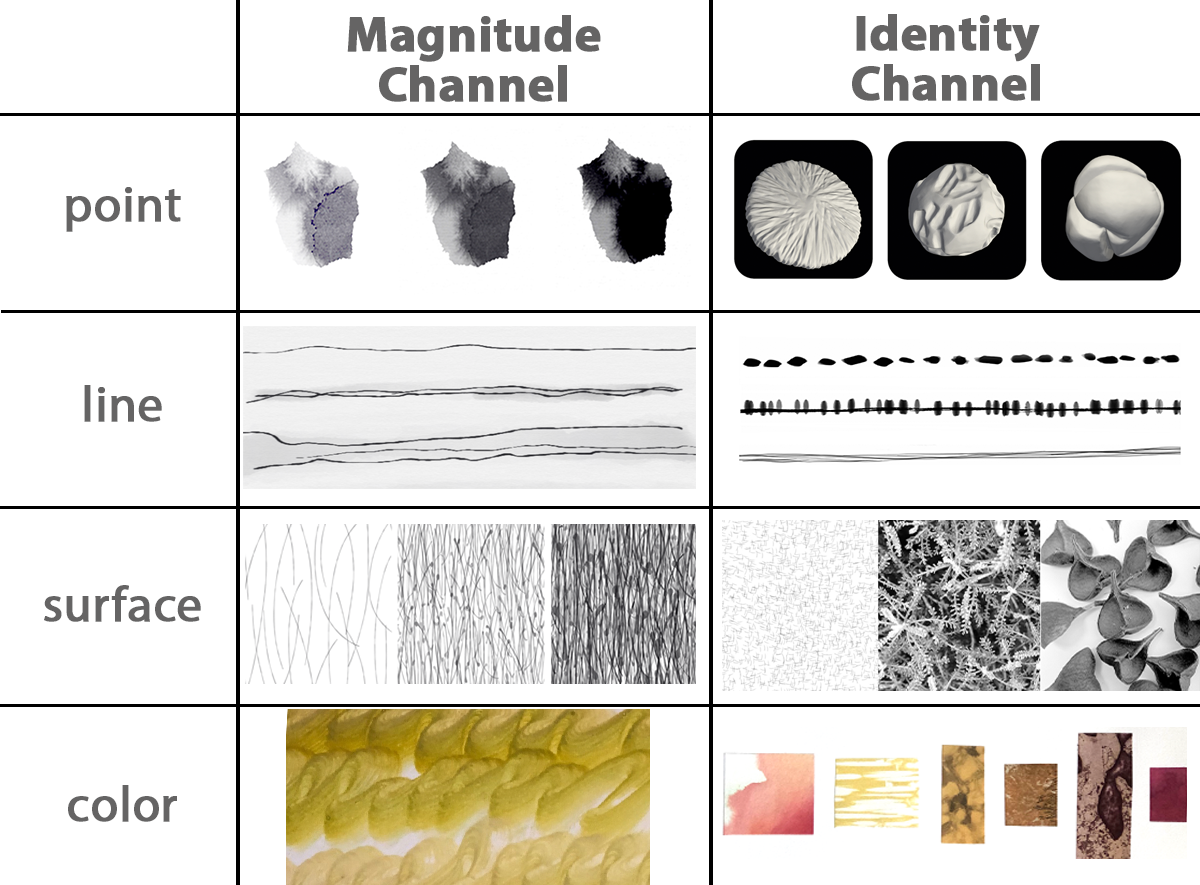}
\caption{Artists will recognize the formal properties of (point, line, form, texture, and color) in these visual examples.  Visualization scientists will recognize magnitude channels to encode ordered data and identify channels to encode categorical data.  The volume category focuses on color schemes for volume rendering algorithms.}
\label{fig:designchart}
\end{figure}

\subsection{Stage 1: Creating and Curating Artifacts}
Stage 1 of ABR is concerned with making or curating \textbf{artifacts}, physical representations of color, line, texture, or form that are the elemental visual building blocks of the final visualizations.  Artists are used to thinking in these terms.  Artists build, analyze, and deconstruct visual scenes using design elements known as {\em formal properties}: point, line, shape, form, texture and color~\cite{ElementsDesign,DesignBasics}.

Recognizing the similarity in the way artists define formal properties and the way scientific visualization practitioners characterize the topology of underlying data variables (point, line, surface, volume), we organize artifacts as diagrammed in Fig.~\ref{fig:designchart}.  Notice that artifacts are grouped not just by data topology, but also by use, following Munzner's classification~\cite{munzner2014visualization} that distinguishes between visual marks for effectively encoding ``magnitude'' relationships (e.g., scalar temperature data) and marks for encoding ``identity'' relationships (e.g., phytoplankton vs zooplankton).  

Artifacts may be sculpted with 3D artistic tools; we have experimented with clay sculpturettes, imprints, and shaved wax.  2D artifacts are also useful, and we have experimented with drawing, painting, texture rubbings, prints, and photography.  Finally, artifacts can be acquired from an endless number of found objects, and these can be arranged in patterns to produce even more vis assets.  We have worked with leaves, rice, lentils, rice noodles, seed pods, gravel, photographs of friends' paintings, and found photographs.  Color artifacts can come from many sources, including painting and collage.

\subsection{Stage 2: Digitizing, and Translating Artifacts}


\begin{figure}[tb]
\centering
\includegraphics[width=\columnwidth]{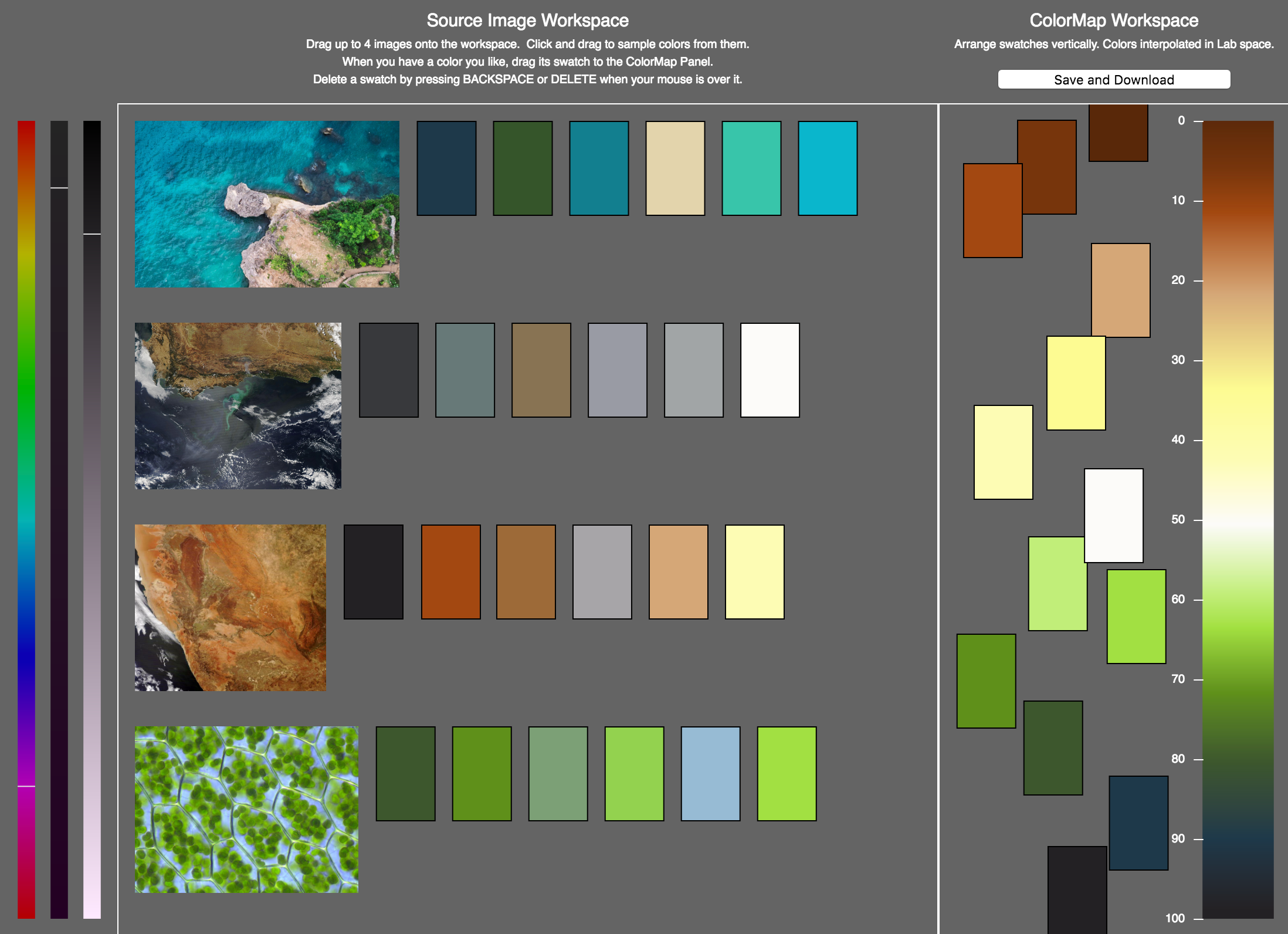}
\caption{The Color Loom applet. Artists drag and drop source images into the left panel and pull swatches of color from these, which are then copied and arranged in the right panel to create a color map.}
\label{fig:AppletColor}
\end{figure}

\begin{figure}[tb]
\centering
\includegraphics[width=\columnwidth]{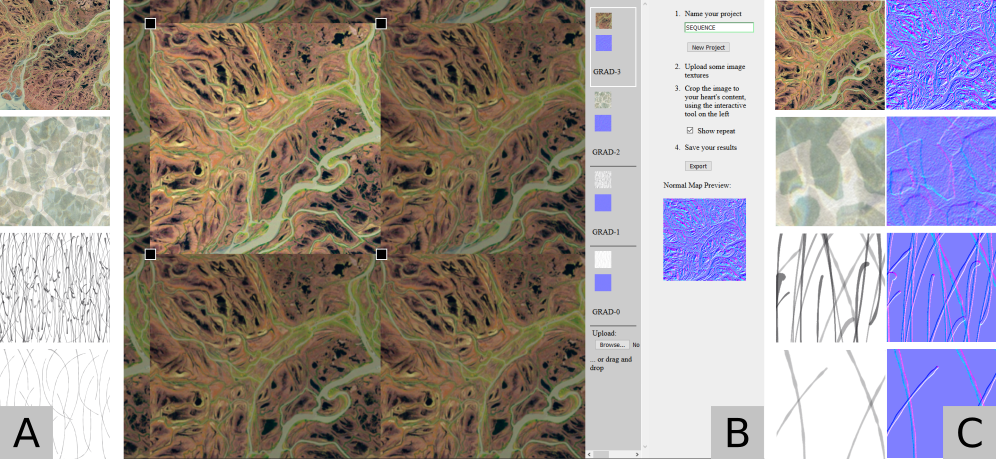}
\caption{The Texture Shaper applet.  A: Original source images, B: Selecting a cropping box; C: Output images and normal maps.}
\label{fig:AppletTexture}
\end{figure}

\begin{figure}[tb]
\centering
\includegraphics[width=\columnwidth]{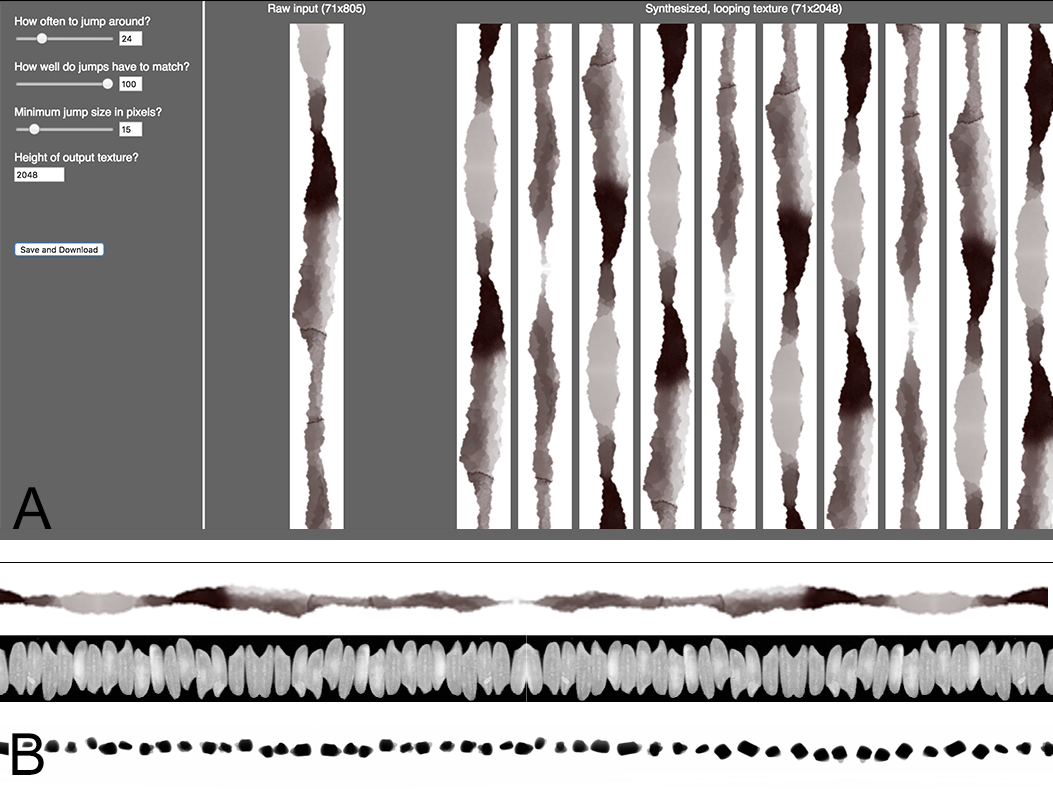}
\caption{The Infinite Line applet.  A: The user interface with parameter controls and texture synthesis preview.  B: Examples of textures synthesized from inkwash, rice grains, and ink dots.}
\label{fig:AppletLine}
\end{figure}

Stage 2 of ABR begins with capturing the material appearance and/or form of the physical artifacts produced in Stage 1 as \textbf{digitized artifacts}.  Here, the specific capture technique depends completely on the type of artifact and sometimes also on the intended eventual use.  Next, digitized artifacts are translated into \textbf{vis assets}, for example, color maps defined in a standard color space, computer graphics-ready textures, glyph meshes that are correctly oriented, down-sampled if necessary, and have normal maps applied for fast rendering.

\subsubsection{Digitizing Material Appearance and Form}
A variety of capture techniques can be used, and the current implementation demonstrates several options for both 2D and 3D artifacts.  Photography (e.g., digital photographs of seed pods and bark found in nature), scanning (e.g., scans of hand-painted ink wash lines or painted color maps), or digital tools with physical inputs (e.g., drawing boards) are used to capture handcrafted 2D material appearances.  In the future, we are keen to also include lighting dependent material appearances using methods, such as linear light source reflectometry~\cite{gardner2003linear}.

Photogrammetry and structured light 3D scanners are used to capture 3D material appearance and form.  Most of the digital artifacts pictured here were captured using an automated EinScan-SE structured light 3D scanner.  Each scan produces a high-resolution mesh of around 100,000 vertices with corresponding photographic texture data. In the future, low-cost smartphone based scanning might also be used (e.g.,~\cite{muratov20163dcapture}).

An open online digital library and underlying database system stores the raw digitized artifacts and the vis assets that are generated from them in the next step.  Metadata classifying the artifacts based on material type and possible use (e.g., to encode line, direction, points) are included to enable online searching and filtering.

\subsubsection{Translating to Vis Assets}
Digitized artifacts are translated to vis assets using four custom interactive applets.  These are needed for two reasons.  First, processing is often required before raw scans can be used in the rendering pipeline.  Second, we wish to be able to reinterpret each raw digitized artifact in multiple ways as a vis asset (e.g., a scanned 3D mesh might be used both to define a 3D glyph shape {\em and} a normal map for a bumpy texture to apply to an isosurface).  User interactions with the four applets are demonstrated in the accompanying video.

\textbf{Applet 1: Color Loom.}  The Color Loom applet (Fig.~\ref{fig:AppletColor}) helps artists to weave colors from digitized photographs, paintings, and other artifacts together into coherent colormaps.  First, the artist drags-and-drops one or more source images onto the left panel of the window. A modified median cut quantization algorithm~\cite{heckbert1982color,bloomberg-2008-modifiedmeancut} then identifies a suggested palette of six prominent colors found in the image, which are displayed as editable color swatches to the right of the source image.  Artists can manually pull additional colors from the source images and create more swatches by hovering over specific pixels.  The color of any swatch may be tweaked using hue, saturation, and brightness sliders.  To build a colormap, swatches are dragged to the right panel, where their vertical positions define control points for a continuous colormap with interpolation performed in CIE Lab space. Results are saved in Paraview .XML and PNG image formats.

\textbf{Applet 2: Texture Shaper.} The Texture Shaper applet (Fig.~\ref{fig:AppletTexture}) supports cropping, previewing repeating texture patterns, and saving results in a standard format expected by the ABR rendering engine.  It also supports more advanced features that are useful for ABR, including converting source imagery into normal maps that can be used in per-pixel lighting calculations for 3D rendering and building ordered texture sets for encoding data, like a binned gradient.  Results are exported from the applet as (sets) of compressed PNG image(s).

\textbf{Applet 3: Infinite Line.} The Infinite Line applet (Fig.~\ref{fig:AppletLine}) uses texture synthesis to transform an example image of a vertical linear mark into a longer, seamlessly repeating texture that can be mapped onto ribbons and other 3D forms.  The algorithm follows the ``video textures'' algorithm presented by Sch\"odl et al.~\cite{schodl2000video}, which  has been used to synthesize drawn and painted strokes for computer graphics non-photorealistic rendering~\cite{kalnins2002wysiwyg} and is a natural fit since only 1D texture synthesis is required.  A similarity measure is computed, comparing each row of the texture to every other row.  Then, a new texture is synthesized, starting from a random starting row and proceeding through the texture by either moving to the next row in the original source or, with some probability, jumping to a new similar row.  A heuristic is used to make the final image loop seamlessly; after synthesizing an image five times larger than the desired output height (typically 2048 pixels), the algorithm searches through the result to find the subsection of the desired height where the starting and ending rows are most similar.

Since outputs of texture synthesis algorithms like this one are highly dependent upon the algorithm's parameters (e.g., probability of ``jumping'' to a new row, the minimum allowable quality for ``jumps'', the minimum size of a ``jump''), the applet makes it possible for artists to adjust these parameters and view the results in real time, preserving the visual characteristics that encode identity while avoiding the distracting regular pattern that is visible with a regular, repeating tiled texture.

\begin{figure}[tb]
\centering
\includegraphics[width=\columnwidth]{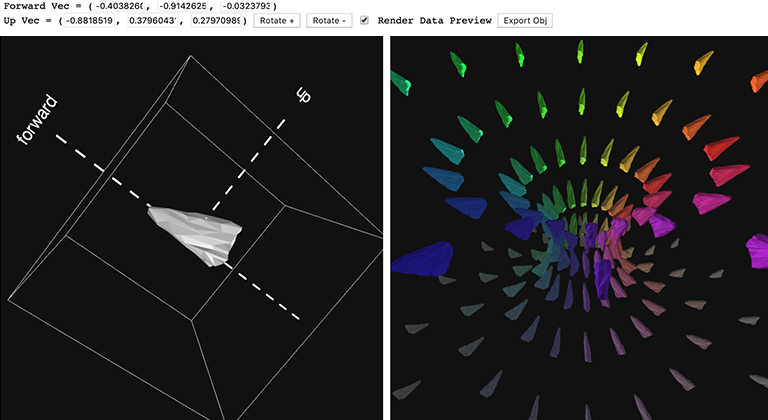}
\caption{The Glyph Aligner applet. Artists use trackball controls in the left panel to align a 3D scanned glyph. The right panel provides a glyph field preview using synthetic data.}
\label{fig:AppletGlyph}
\end{figure}

\textbf{Applet 4: Glyph Aligner.} The Glyph Aligner applet (Fig.~\ref{fig:AppletGlyph}) works with 3D scanned artifacts, which require user input and data processing before being used as vis assets.  On the left, mouse-based camera and object trackball controls are used to reorient the glyph to associate ``forward'' and ``up'' directions with the 3D scan.  On the right, a preview updates in real-time to show the result of applying the glyph to visualize an example vector field.
\label{sec:blender}
After interactively aligning the glyph, the resulting mesh is passed to an automated Blender3D Python script.  This script decimates the 3D scanned mesh, which may include 100,000+ vertices, to varying degrees to support level-of-detail (LOD) rendering in VR, while preserving detail with normal mapping. A UV mapping for each mesh is creating using Blender's ``Smart UV Project'' algorithm, and normals are stored based on the original geometry.  Then, for each LOD mesh the differences between these original normals and the normals of the decimated mesh are baked into a LOD-specific normal map.  The output can reduce the vertex count by three orders of magnitude while preserving much of the surface appearance.

\subsection{Stage 3: Data-Visual Mapping and  Visualization}

Stage 3 of ABR involves implementing data-driven interactive visualizations using the vis assets.  To provide structure to these multivariate visualizations, we say that each visualization is composed of multiple \textbf{vis layers}.  These are analogous to the 2D layers used by artists in image editing programs, but vis layers are not 2D; they are true 3D volumetric constructs.  Each vis layer has hooks for connecting vis assets to \textbf{data objects}.  Thus, after creating a new oriented glyph layer, a designer could attach vis assets to define the 3D glyph mesh and color map to use for the vis layer and also attach data objects (e.g., density sampled phytoplankton concentration, velocity magnitude) to drive color changes.  The layers are combined into a final \textbf{interactive visualization}, which may be rendered fast enough to optionally display in head-tracked, stereoscopic VR.

\subsubsection{Defining a Spec. for an ABR Rendering Engine}

We begin by more completely defining the problem an ABR rendering engine must solve.  An ABR rendering engine must produce 3D computer graphics imagery of various data topologies (points, lines, surfaces, and volumes) using, as directly as possible, the visual styles defined by real-world physical artifacts.  This is a hard problem because it requires a balance between staying true to the original visual properties of the artifacts and manipulating these based on underlying data.  Also, artifacts can be interpreted in so many different ways.  Given a series of evocative, organic textures captured from leaves, rocks, or seeds, how precisely should they be used to encode data categories or magnitudes?  The answer will likely change based on the data topology (points, lines, surfaces) and on the goals of the visualization.  Thus, we reason that, like all good design tools, a good ABR rendering engine should provide options (multiple complementary rendering techniques) for how to interpret artifacts and attach them to data.  We think of the possible techniques as existing along a spectrum. 
\begin{figure}[h!]
\centering
\includegraphics[width=\columnwidth]{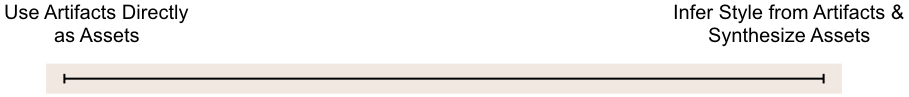}
\end{figure}

On the left are ABR rendering techniques that can be implemented with little or no change to traditional visualization rendering.  Color is one example.  Existing systems typically support rendering a constant color per data object for identity encodings, or color mapping for magnitude encodings.  If the colors come from real-world source images, this can be considered an ABR rendering technique, albeit at the far left of the spectrum.

\begin{figure*}
\centering
\includegraphics[width=0.9\linewidth]{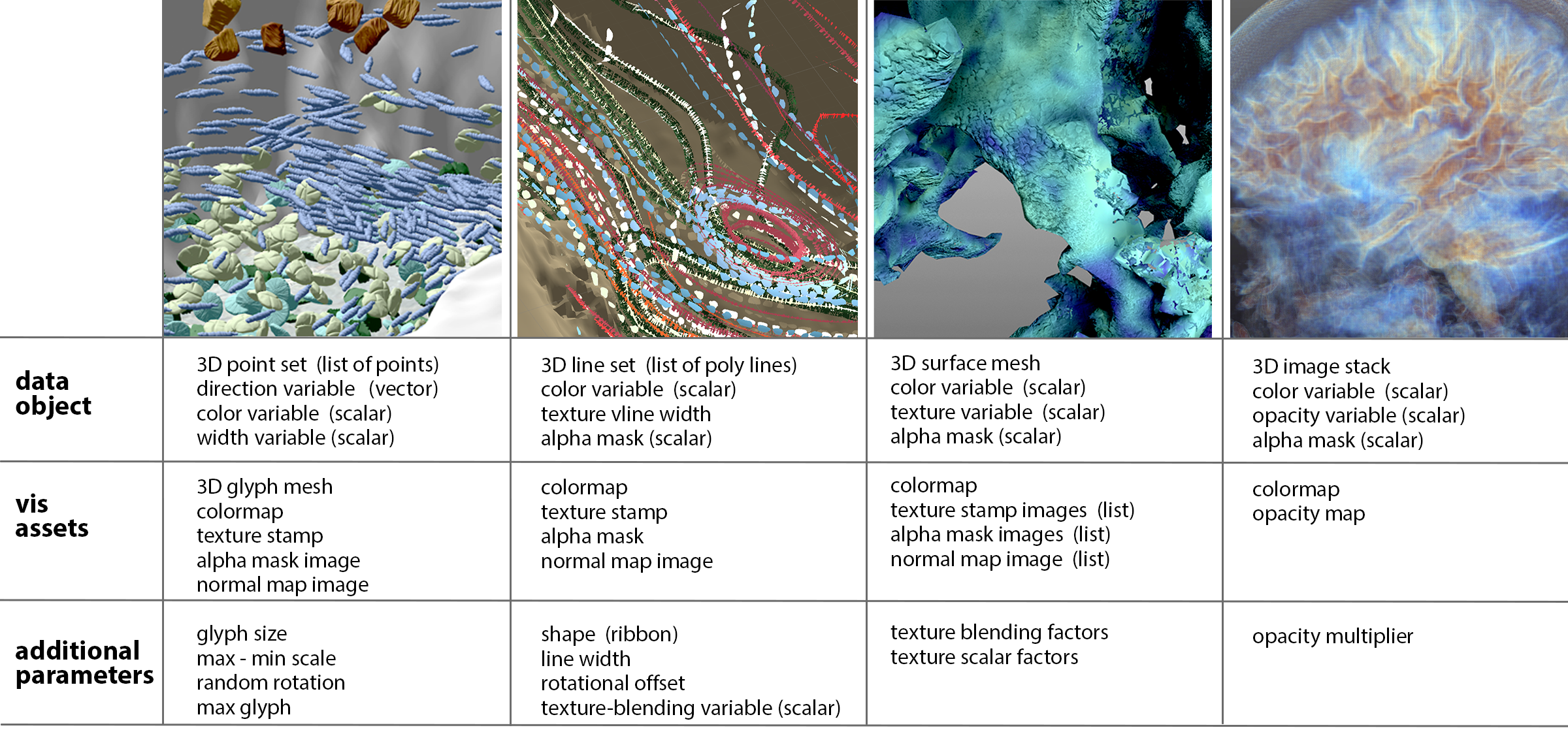}
\caption{Example renderings and parameters for vis layers in the ABR rendering engine.}
\label{fig:layerTypes}
\end{figure*}

Beyond color, most ABR rendering techniques are non-trivial to implement in available visualization rendering engines.  For example, existing tools often limit 3D glyphs to preset geometric primitives.  Textured lines or surfaces are not difficult to implement from a computer graphics standpoint, but scientific visualization engines do not typically expose an ability to set the texture of lines and surfaces in order to support identity encodings, and we know of no existing 3D scientific visualization design tools that support varying these textures to support magnitude encodings.  We demonstrate each of these ``middle of the spectrum'' techniques.

On the right of the spectrum, we predict specific ABR rendering techniques may themselves be the subject of future computer graphics research.  For example, given a set of clay glyphs used to encode twist along a line, computer vision and machine-learning algorithms could be used to infer to visual style implied by the artists' examples and synthesize new parametrically controllable forms.  The Infinite Line applet described earlier is an important step toward this goal of automated synthesis of vis assets from artifact examples.

ABR rendering techniques must support both identity channel and magnitude channel encodings (see Fig.~\ref{fig:designchart}).  Identity encodings are generally easier.  Magnitude encodings are more innovative, and there are at least three possible approaches, ordered moving from left to right along the spectrum: (1) Mapping an ordered set of artifacts piecewise to data, like a binned color map. (2) Using artifacts as control points, implementing some form of morphing between them, and mapping the results to continuous data. (3) Parameterizing some characteristic of an artifact and synthesizing new instances with data controlling this parameter.  This could be as simple as controlling the width of a glyph artifact, or as complex as synthesizing new texture patches with higher density distributions of a pattern in response to underlying data.

\subsubsection{Mapping Data to Visuals in Layers}

Building upon metaphors that work for artists in other visual design tools (e.g., image editors), we think of each new visual addition to the scene as a layer.  Every vis layer takes as input one data object, some number of scalar or vector variables of that data object, and some number of vis assets, all of which may be selected at design-time by the user (Fig.~\ref{fig:layerTypes}).  The extensible rendering engine currently supports four layers.

\textbf{Glyph Vis Layer.} The glyph vis layer renders instances of a glyph (either a 3D mesh or an image displayed on data-aligned quads) located at coordinates specified by a point set data object.  The glyphs can be axis-aligned, aligned to a selected direction vector variable, or assigned random orientations.  Color can be assigned either as a constant across all glyphs or data-driven. A constant glyph size can be selected (defined as a percentage of the largest extent of the data object), and an axial radius can be specified either as a constant, or according to a selected width scalar variable. Normal maps are automatically applied if available. For image-based glyphs, alpha masks may be supplied to mask away the negative space.

We found that an important use for custom-sculpted glyphs is to represent scalar field data by distributing glyphs on a surface or throughout a volume.  Thus, regular, random, and density-based samplings are supported.  For density-based sampling, we implemented a Metropolis-Hastings algorithm~\cite{chib1995understanding}, which is a Markov Chain Monte Carlo method.

\begin{figure}
\centering
\includegraphics[width=\linewidth]{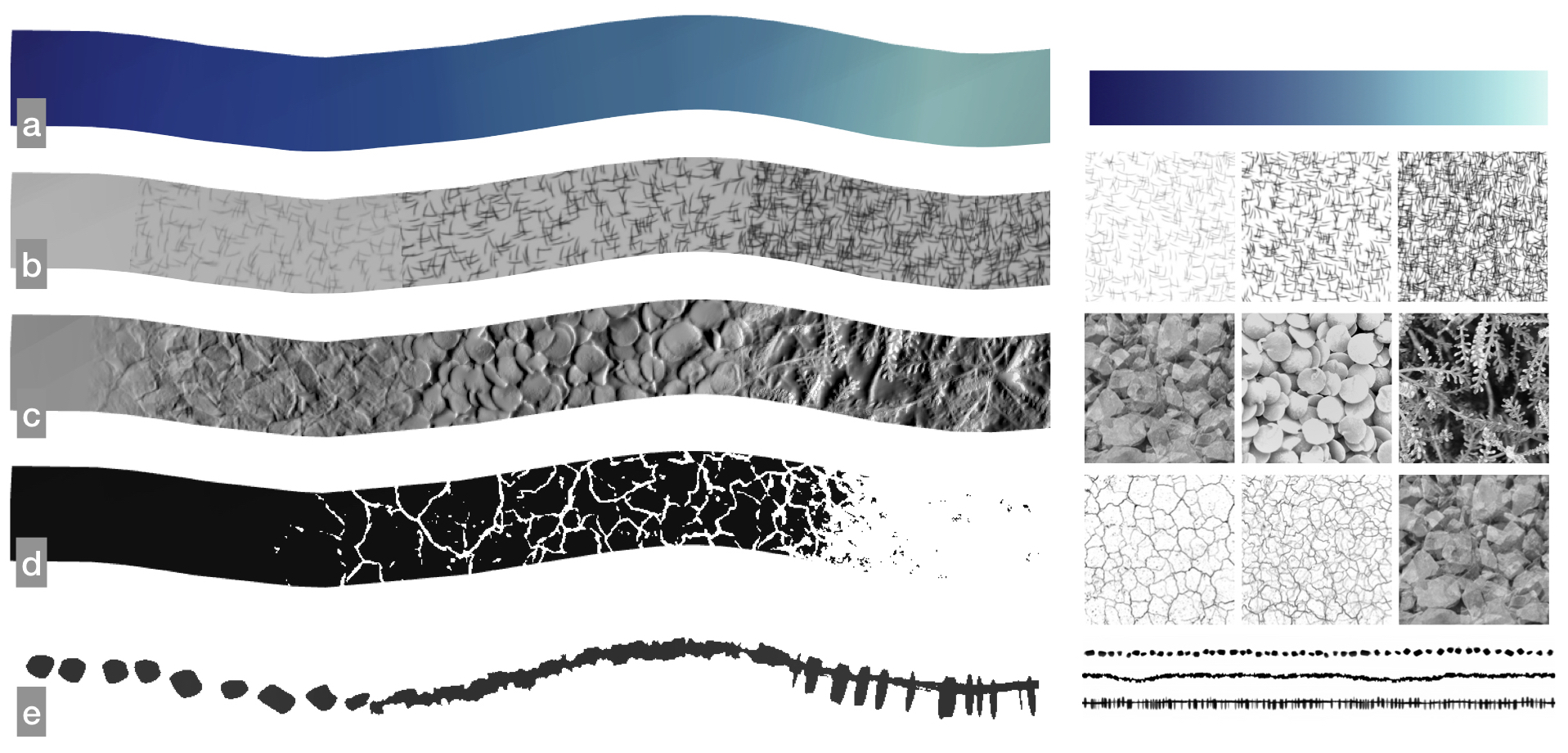}
\caption{Options for applying vis assets to lines or surfaces include (a) color mapping, (b) data-driven texturing, (c) data-driven texturing with bump mapping, (d) data-driven texturing with blending and masking, (e) data-driven texturing with masking to create an organic line profile. }
\label{fig:linetyles}
\end{figure}

\begin{figure*}[tb]
\centering
\includegraphics[width=\textwidth]{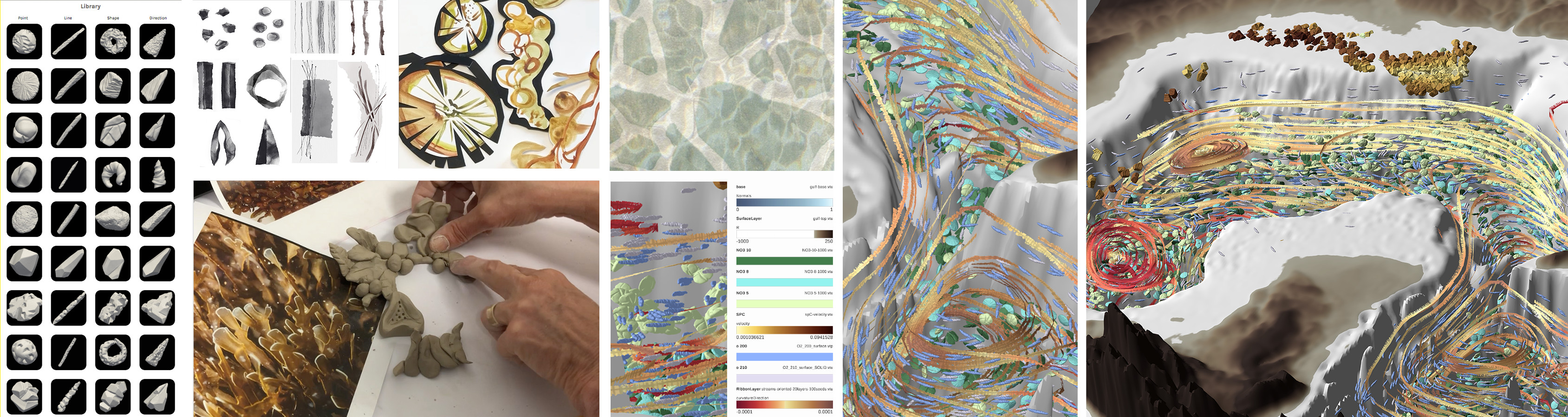}
\caption{Process and results from the internal exploratory design study with ABR on the biogeochemistry data in the Gulf of Mexico \cite{wolfram2015diagnosing}, left to right: pre-made glyphs; glyphs painted during the study; glyphs constructed during the study; textures captured pre-study; detail of final visualization; visualization of the Gulf of Mexico.}
\label{fig:study-ABR}
\end{figure*}

\begin{figure}[tb]
\centering
\includegraphics[width=\columnwidth]{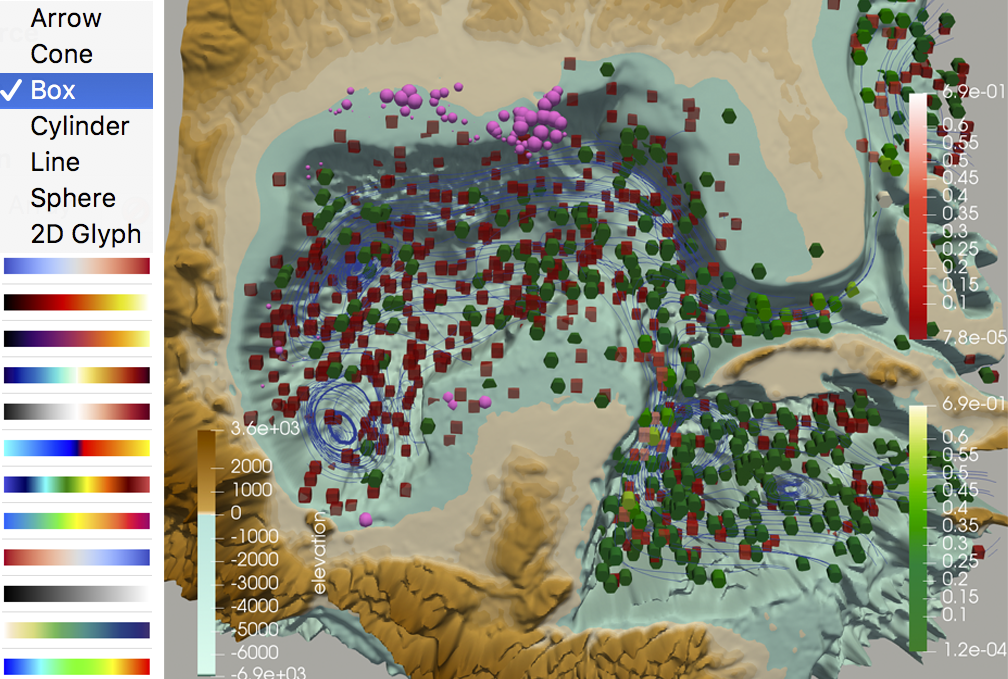}
\caption{Left - Encoding options in ParaView, Right - Results from the internal exploratory design study with Paraview.}
\label{fig:study-PV}
\end{figure}

\textbf{Line Vis Layer.}  The line vis layer renders ribbons or tubes along paths defined by a line set data object.  If normal vectors for each point are provided, the ribbon or tube can be oriented according to the line normal and some rotational offset, and each mesh vertex is assigned texture coordinates with $u$ equaling the arc-length from the line origin to the vertex, and $v$ running from 0 to 1 either across the ribbon, or along the circumference of each tube ring. Color is assigned either as a constant along the entire geometry or based on a data-driven colormap.  Similarly, a single tiled stamp, alpha-mask, or normal map can be applied across the entire geometry.

Optionally, an ordered texture set can be used for data-driven texturing.  In this mode, textures are applied using a binned data mapping; the data range is divided into $N$ evenly-sized bins, where $N$ is the number of textures in the corresponding texture set.  Blending can be applied in the fragment shader to hide texture seams, with a user-defined blend distance.  Fig.~\ref{fig:linetyles}b-e shows specific examples of how textures are binned and blended. 

Our implementation supports multiple sampling strategies and artists can switch between samplings that have equal steps in arc length or integration time, making it possible for the visual style to include evenly placed glyphs that are distorted (stretched) to encode speed or to include undistorted glyphs that are spaced along the line to encode speed.

\textbf{Surface Vis Layer.} The surface vis layer renders a triangle mesh defined by a polygonal mesh data object.  Every fragment of the surface can be colored and textured based on either constant vis assets, or vis assets that are blended by the provided data variables.  Since the surfaces are often complex and have no inherent UV texture parameterization, an automatic texture mapping approach is needed.  We implemented a tri-planar projection technique, where the texture is projected along the axes and the three projections are blended according to the normal of the surface at any given location.  This blending is further influenced by a projection blending factor that controls how crisp the seams are between the three projections.  If the texture has a clear structure, such as grains of rice, a low blending factor may be preferred, whereas if the texture is more continuously varying, such as watercolor strokes, a higher blend factor will help hide the seams.  
All of line-style effects shown in Fig.~\ref{fig:linetyles}a-d can also be applied to surface layers to do data-driven texturing on arbitrary 3D surfaces.

\textbf{Volume Vis Layer.} Artists often refer to density fields of glyphs created with Glyph Layers as a ``volumetric effect''.  However, the engine also supports visualizing volumetric scalar fields using traditional volume rendering.  The volume vis layer volume-renders a 3D grid of voxel data using artifact-based colormaps as a transfer function.

\subsubsection{User Interface and Implementation}

Although it may be possible to implement ABR on top of a variety of other existing rendering pipelines using the concepts and framework described here, before this work, there was no existing system that could support the rendering required without a significant system-building effort.  Our implementation is built upon a combination of VTK, for access to advanced data processing routines, and Unity, for better support for rendering in a variety of immersive displays and future support for 3D and tangible user interface techniques.  All of the software, artifacts, and assets are being made openly available to the public.

The Unity user interface for creating and defining vis layers is generated dynamically at runtime based on the installed vis layers.  It includes an ``Add New Layer'' functionality, where the user selects the layer type from a list.  Each layer then creates a new panel in the interface where each parameter of the layer may be edited.  Data object parameters are also defined dynamically based on the VTK objects and variables currently loaded and include a data-range interface to interactively specify min and max data values during design.

\begin{figure*}[tbh]
\centering
\includegraphics[width=\textwidth]{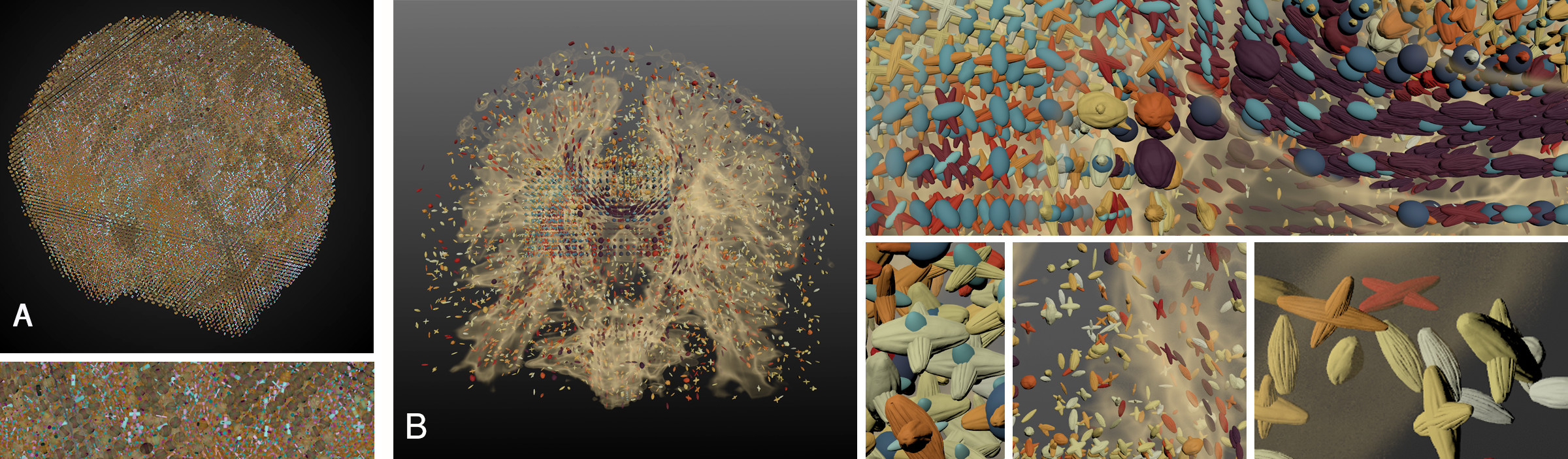}
\caption{Visualizing brain microstructure in 3D.  A: Straightforward extension on slice-based methods to 3D. B: ABR designed visualization with high-resolution data in the centrum semiovale.}
\label{fig:Brain}
\end{figure*}

\section{Internal Exploratory Design Study}

The quality of the work that results from design processes is one possible metric for evaluating ABR, and we investigate this with domain science collaborators in Section~\ref{sec:applications}.  However, having followed the proceedings of the BELIV workshop series that explores visualization evaluation techniques ``beyond time and errors''~\cite{beliv2018}, we were motivated to consider more creative alternatives for evaluating impact on the visualization design more directly.  Thus, the work in this section targets more direct metrics, such as the number and quality of design alternatives explored when faced with a real-world, challenging scientific visualization design problem.

Our long-term goal is to measure this type of impact on process in a larger-scale setting (e.g., building on recent workshops at IEEE VIS~\cite{discoveryjam1,discoveryjam2} and the College Art Association Conference~\cite{meetthescientists}).  The 2-day internal exploratory design study presented here is intended to be a first logical step toward this, providing an initial characterization and comparison of the design process with ABR vs Paraview, which serves as an example of a popular traditional scientific visualization tool, and piloting possible future evaluation methods and metrics.

\subsection{Methodology}

Given a new scientific dataset and time to discuss relevant data and research questions with a domain scientist, the \textbf{task} is to explore a variety of potentially useful visualizations for the problem.  Since the design study is posed as an $A$ vs $B$ comparison, the artist performs that same visual design task first with ABR and then with Paraview.  Learning effects are sure to be a factor, but, by scheduling ABR on day 1 and Paraview on day 2, the advantage falls to Paraview.  

Our research team's artist, also the second author of the paper, acted as the only \textbf{participant} for the internal design study.  Being a member of the research team, the artist had worked to co-develop ABR for more than 18 months.  However, all of the features of the system were not ready until the week before the design study.  Thus, her hands-on \textbf{training} creating art with traditional materials was a lifetime; her hands-on experience with the ABR toolset was less than 1 week; and her hands-on experience with Paraview was more than 5 years.

In a true workshop setting a formal introduction to the data would be required, but our case, the artist and scientist had already discussed high-level scientific goals as part of our ongoing collaborative project.  The \textbf{dataset} comes from scientists studying mariculture, specifically commercially viable macroalgae growth.  They have a challenging data analysis problem that requires understanding of many critical data variables from a fusion of remote sensing and a high-resolution computational simulation of ocean currents~\cite{ringler2013multi,petersen2015evaluation,wolfram2015diagnosing} extended to include biogeochemistry \cite{moore2001intermediate,moore2013marine,wang2014impacts,wang2015influence}.  In offshore regions of the Gulf of Mexico, eddies in the ocean currents could provide ``small farms'' for macroalgae since the eddies can carry key nutrients, but to understand which eddies provide the most suitable conditions, it is necessary to relate eddy velocity, rotation, divergence, salinity, and temperature data together with nitrate, phosphorus, and other biogeochemical variables, a few of the 30 variables within the dataset -- all within the context of the local geography.  To facilitate direct comparison, the same data were utilized on day 1 and day 2.

We \textbf{recorded the design process} by saving state files and screenshots at regular intervals and whenever an interesting image was produced.  We also logged time performing design functions (e.g., refining color palettes, sculpting glyphs, sketching).

\subsection{Results and Interpretation}

Figs.~\ref{fig:study-ABR} and \ref{fig:study-PV} document process and results for the two tools. Since rapid experimentation and broad thinking is the foundation of all creative design processes~\cite{buxton2010sketching}, we also report data on the number of designs and design elements explored and time performing design activities.  With ABR the artist designed with 72 line textures and glyph artifacts.  49 artifacts were created during the study, including 43 new hand-painted artifacts; 6 new hand-sculpted glyphs and 16 new colormaps. 

The total time working with ABR was 7 hours, 46 minutes.  2 hours, 19 minutes were spent on Stage 1 of the ABR pipeline (crafting and making), with 8 minutes of that time devoted to searching for reference imagery online and the rest working with clay, paint, and wire.  The total time spent on Stage 2 (digitizing and translating) was 3 hours, 25 minutes.  The total time spent on Stage 3 (data-visual mapping and VR visualization) was 2 hours, 2 minutes. 

With Paraview, 4 existing glyphs, testing a range of sizes, were explored from the 5 available built-in geometric primitives, and 14 existing colomaps (Fig.~\ref{fig:study-PV}.  The time working with Paraview was shorter than expected going into the study; just 2 hours and 43 minutes.

The most interesting result is the difference in time and motivation devoted to design with each tool.  Paraview includes just three non-directional glyphs, so the options for glyph combinations were quickly exhausted. If the artist had not already spent time exploring color on day one, more time could have been spent usefully exploring color in Paraview -- this is the area in which artists are most able to contribute to design with Paraview. However, since she had already spent considerable time on color, she quickly reached a point of diminishing returns on day two; frustration due to limited options for glyph forms grew once she had balanced the size and color intensities of the nitrates, and she did not want to continue.  

By contrast, with ABR, the artist worked over 7.5 hours to test the basic glyph and line design options. She reported that the power and versatility of ABR gave her a range of design options comparable to those available in her physical studio, where possibilities are essentially limitless. A key finding was that though it may seem that ABR would be a comparatively slow method of designing glyphs, in fact, the artist was able to iterate through many forms based on the needs of scientists with more speed and more precise results than she was able to with previously available tools. This mode of quick, preliminary iteration and experimentation was crucial in harnessing the power of the increased visual vocabulary that ABR enables. The process resulted in a broad range of possible solutions suitable to addressing the complex visualization problems posed by large, multivariate datasets and complicated scientific questions. While making individual forms was not in itself time-consuming, with such a capacious range of options for design, the artist did need to spend more time honing in on those best suited to the research questions and visualization needs.

Qualitatively and quantitatively, we can say that ABR enabled a broader exploration of the visualization design space.  Shneiderman outlines requirements for computer-based creativity support tools as enabling results that are both novel and useful~\cite{schneiderman2007creativity}. Simply comparing the process and result imagery in Fig.~\ref{fig:study-ABR} to typical VIS proceedings makes a clear case for novelty.  Despite the limitations of this early study (a single impossible-to-be-unbiased user, an exploratory rather than scientifically controlled study), we believe these results already demonstrate that ABR is enabling us to reach and visually critique points in the scientific visualization design space that we have never explored before.  The study does not specifically address usefulness.  Thus, we take an early step toward this evaluation in the next section.

\section{Applications and Guidelines}
\label{sec:applications}

We have also evaluated ABR by applying the new framework together with collaborators on two actively researched scientific datasets.

\subsection{Macroalgae in the Gulf of Mexico}

The first application uses the same Gulf of Mexico data as the design study.  This section reports on follow-on efforts with these data, taking a week of time to refine the visualization and seek feedback.

Fig.~\ref{fig:teaser} along with the accompanying video document the visualization results achieved using ABR.  Color maps were generated to provide a natural color palette using artifacts from photographs and paintings.  Eddies are shown using textured ribbons to depict the flow lines.  Color encodes rotational direction (green-blue for cyclone, orange-red for anti-cyclone), and the texture itself is varied along each ribbon based upon the local degree of curvature.  Temperature and salinity are encoded using textured surface layers.  Three isosurfaces of temperature are shown at levels to evaluate macroalgae growth (20, 25, and 26 degrees Celsius).  These are also colored and textured using an inkwash texture set; the texture is denser when salinity increases.  Finally, three custom glyphs are distributed using density based sampling to depict three types of nitrates.  The colors for the three were chosen to be analogous and also vary in response to the local salinity.

The collaborating domain scientist posed the challenge of being able to clearly visualize 5 or more variables simultaneously because he has found this impossible to accomplish using his current toolset consisting of Paraview (via slices, isosurfaces, and volume rendering) and python-matplotlib (to render final plots for publication after finding areas of interest via Paraview).  Our interpretation is that these current tools are not visually expressive enough -- one can only overlay so many surfaces and volume clouds before the different fields become too difficult to discern.  Upon seeing the results in Fig.~\ref{fig:teaser}, the scientist reported that ABR will be ``transformational'' to his science, saying these pictures can easily visualize more than five variables.  With this type of glyph-based visualization, ``you can superimpose functional relationships between multiple fields [such as nitrates getting caught in eddies and pulled to the surface] in a way that you can't with volume rendering or surfaces alone.''  Similarly the textured isosurfaces with color have the potential to encode ``three in one'', packing more data into the multivariate visualization.  On the aesthetic, he commented: {\em They look like biology more than they look like plastic.  They could be real, produced by nature. I think that people are going to underestimate that.  At first, I'm perturbed, these don't look like plastic, then I realize this is not a problem but a major benefit.}

\subsection{Brain Microstructure Imaging}

In another application domain, scientists are developing computational tools to leverage high-field Magnetic Resonance Imaging (MRI) for understanding structural and functional alterations of brain connections in neurodegenerative disorders.  The latest algorithms in this field make it possible to not only identify several crossing pathways in white matter areas with complex configurations but also to estimate microstructural parameters, such as axonal diameter and density~\cite{farooq2016brain,farooq2016microstructure}, increasing the difficulty of visualizing Diffusion-Tensor MRI data, which is already regarded as a significant 3D visualization research challenge.  To date, such data have only been visualized using slice-based approaches (e.g.,~\cite{farooq2016microstructure}).

Fig.~\ref{fig:Brain}A is a straightforward translation of the prior slice-based visualizations to a true 3D visualization.  Clearly occlusion is a major factor in designing an effective 3D representation, and this straightforward translation is not successful.  Section B shows a refined ``magic lens'' design.  Volume rendering and sparsely sampled glyphs provide context throughout the brain, and high-resolution data are presented in an interactively defined small volume.  The voxels are filtered to display only regions with crossing fibers using oriented glyphs to show the primary and secondary fiber orientations.  A glyph set was designed so that each glyph has a similar profile, but the density of the linear texture along the length of each glyph increases in response to the axonal density variable.  The glyphs are also sized based on the axonal radius parameter.  An oriented ellipsoid depicts the ``leftover'' diffusion for each voxel after computing the two primary fiber orientations, and a volume rendering provides anatomical context.

The collaborating domain scientist suggested a focus on the centrum semiovale (highlighted in Fig.~\ref{fig:Brain}B), which is a region with high fiber crossings.  From the primary and secondary fiber orientations, the visualization confirms expected brain structure in this region. This is the first time the scientist had seen the data in a true 3D display, and looking at the visualization in VR convinced him that he can see structure that is not possible to see in 2D slice-based visualizations.  Similar to the Gulf of Mexico results, the aesthetic produced with ABR leads to a natural, organic visual language, and by encoding data with textural variations of the glyph rather than color (as used in prior slice-based approaches), we were able to use color to encode a new derived variable (similarity between primary and secondary directions), which, with a color map applied, calls visual attention to crossings that are nearly perpendicular (red color range).

\section{Discussion}

\subsection{Abstract Data and Future Work}
There are many other possible applications that remain to be tested, such as more abstract data that do not exist within a predefined 3D structure.  We believe the approach will translate well to abstract data because artists design visuals to convey abstract concepts, like human emotion, all of the time.  It would be fascinating, for example, to ask artists to craft a series of glyphs to represent generic uncertainty, correlation and anti-correlation, or cause and effect.  Future work also includes experimentation with new vis layers suggested by artist users and combining the handcrafted, organic aesthetic that is possible with ABR with more traditional geometric aesthetics. 

\subsection{ABR Design Guidelines}

We can provide some preliminary design guidelines informed by the two applications and work with ABR thus far.  At a high level, we note that the structure shown in Fig.~\ref{fig:designchart} provides artists with artifact categories common to their visual language.  Designing a visualization follows the same process one uses when laying out the structure of a painting; blocking out the underlying structure using line, shape and forms. Once the structure is in place, artists can draw upon existing pre-loaded artifacts or create new ones, naturally applying the principles of design (Balance, Repetition/Rhythm, Focal Point/Dominance, Unity/Category~\cite{DesignBasics,ElementsDesign}) to organize the compositions, direct attention, clarify hierarchy, highlight relationships, and create unity.  At a lower level, we have found the following specific design considerations to be most important for working with ABR.

\textbf{Contrasting Forms.} When choosing glyphs, consider contrast between forms. To create a visual contrast, pair glyphs that are: geometric versus handcrafted or organic, curvilinear versus angular, sparsely textured versus densely textured (e.g., the identity channel glyphs in Fig.~\ref{fig:designchart} are arranged from dense to sparse texture).

\textbf{Contrasting Profiles.} Profiles of both glyphs and ribbons play a key role in visual distinction capability. Consider contrast in complexity, profile, and continuity (e.g., the streamlines in Figs.~\ref{fig:teaser} and \ref{fig:study-ABR} (right)).

\textbf{Contrasting Textures.} Consider contrast between size and/or source of surface textures of objects. See  Fig. \ref{fig:study-ABR} (left), for examples of glyphs with contrasting density and source of textures. As with forms, textures follow similar rules in creating contrast; likewise, some examples include organic versus geometric textures and curvilinear versus angular textures as shown in Fig. \ref{fig:linetyles}.

\textbf{Color.} The Color Loom applet enables quick construction of versatile colormaps. When selecting images from which to construct a colormap, select images that provide multiple types of contrast specifically: luminance; cool - warm; and low and high saturation. The range of contrast types will enable high-resolution; semantic associations; and highlighting capability. Ware and Rhyne provide in depth guidance based on these principals \cite{Ware2013visualDesign,rhyne}.


\subsection{Pairing with Perceptual Guidelines}

Aside from giving artist users starting points informed by the perceptual literature, we have made an intentional decision to not constrain artists by encoding hard constraints or rules into the framework, instead favoring giving users the freedom to create and discover new visual encoding strategies that have not been explored before and may break some of the current rules.  If successful, this means that ABR may lead to new visualization designs that researchers wish to verify and better understand using low-level perceptual studies.  Thus, the two approaches are complementary.  ABR might be thought of as a top-down approach to inventing and experimenting with new visual encoding strategies for visualization, whereas low-level perceptual studies of generic uses of color, shape, texture, and form might be thought of as a bottom-up approach.  We are excited for these two  research methodologies to continue to inform each other.

\section{Conclusion}
The internal design study and two applications are by no means a final evaluation; however, they clearly demonstrate an ability for the first time to create complete scientifically useful multivariate VR data visualizations using a visual language derived entirely from traditional physical media.  The resulting aesthetic is novel, looking, as one domain scientist reported, like it could be ``produced by nature''; we believe this has powerful implications for making science more understandable and engaging.  Further, by leveraging skills artists already have with physical media, we believe ABR can have a powerful positive impact on the visualization community by broadening the diversity of people who can now contribute to creating 3D scientific visualizations.


\acknowledgments{
This research was supported in part by the National Science Foundation (IIS-1704604 \& IIS-1704904).  Brain microstructure applications were supported in part by the National Institutes of Health (P41 EB015894, P30 NS076408).  MPAS-O simulations were conducted by Mathew E. Maltrud and Riley X. Brady as part of the Energy Exascale Earth System Model (E3SM) project, funded by the U.S. Department of Energy (DOE), Office of Science, Office of Biological and Environmental Research with analyses conducted by PJW, MEM, and RXB under ARPA-E Funding Opportunity No. DE-FOA-0001726, MARINER Award 17/CJ000/09/01, Pacific Northwest National Laboratory, prime recipient.
}

\bibliographystyle{abbrv-doi}


\end{document}